\documentclass[a4paper]{article}

\usepackage{ISCSLP2024}
\usepackage{colortbl}
\usepackage{xcolor}
\usepackage{stfloats}
\usepackage{ifthen}
\usepackage{cite}

\usepackage[caption=false,font=normalsize,labelfont=sf,textfont=sf]{subfig}
\title{Does Current Deepfake Audio Detection Model Effectively Detect ALM-based Deepfake Audio?}

\name{
	\ifthenelse{\boolean{blind}}{Anonymous to ISCSLP}
	{Name$^1$,Co-author Name$^2$}
}

\name{Yuankun Xie$^1$\textsuperscript{,}$\dagger$, Chenxu Xiong$^2$\textsuperscript{,}$\dagger$, Xiaopeng Wang$^3$\textsuperscript{,}$^4$, Zhiyong Wang$^3$\textsuperscript{,}$^4$,  Yi Lu$^3$\textsuperscript{,}$^4$, Xin Qi$^3$\textsuperscript{,}$^4$, Ruibo Fu$^3$\textsuperscript{,}\textsuperscript{*},Yukun Liu$^4$, Zhengqi Wen$^6$, Jianhua Tao$^5$\textsuperscript{,}$^6$, Guanjun Li$^3$,  Long Ye$^1$}

\address{
  $^1$ State Key Laboratory of Media Convergence and Communication,\\ Communication University of China\\
  $^2$ SDU-ANU Joint Science College, Shandong University, Weihai\\
  $^3$ Institute of Automation, Chinese Academy of Sciences\\
  $^4$ School of Artificial Intelligence, University of Chinese Academy of Sciences \\
  $^5$ Department of Automation, Tsinghua University\\
  $^6$ Beijing National Research Center for Information Science and Technology, Tsinghua University
}
\email{xieyuankun@cuc.edu.cn, ruibo.fu@nlpr.ia.ac.cn}

\begin{document}

\maketitle
\renewcommand\thefootnote{} 
\footnotetext{$\dagger$ denotes equal contribution to this work. * denotes corresponding author.}
\renewcommand\thefootnote{\arabic{footnote}} 
\begin{abstract}
Currently, Audio Language Models (ALMs) are rapidly advancing due to the developments in large language models and audio neural codecs. These ALMs have significantly lowered the barrier to creating deepfake audio, generating highly realistic and diverse types of deepfake audio, which pose severe threats to society. Consequently, effective audio deepfake detection technologies to detect ALM-based audio have become increasingly critical. This paper investigate the effectiveness of current countermeasure (CM) against ALM-based audio. Specifically, we collect 12 types of the latest ALM-based deepfake audio and utilizing the latest CMs to evaluate.  Our findings reveal that the latest codec-trained CM can effectively detect ALM-based audio, achieving 0\% equal error rate under most ALM test conditions, which exceeded our expectations. This indicates promising directions for future research in ALM-based deepfake audio detection.
\end{abstract}

\section{Introduction}

\begin{figure}[t]
	\centering
	\includegraphics[width=\linewidth]{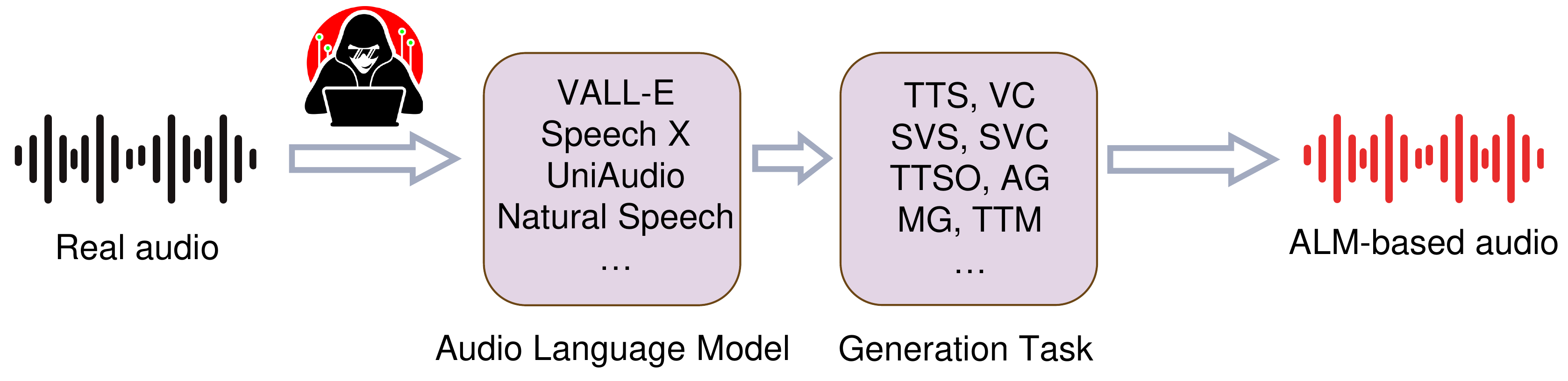}
	\caption{ALM-based deepfake audio generation pipeline}
	\label{fig:almpipeline}
\end{figure}

Currently, due to the rapid development of large language model and audio neural codec, there have been significant advancements in audio generation models. We typically refer to these novel types of audio generation models as Audio Language Models (ALMs) \cite{borsos2023audiolm, kreuk2022audiogen, wang2023neural, agostinelli2023musiclm, zhang2023speak, wang2023viola, copet2024simple, wang2023speechx, chen2023lauragpt, yang2023uniaudio}. These ALM-based audio generation models have lowered the barrier to creating deepfake audio, making the process significantly easier and more accessible. 
On the other hand, ALM-based audio is characterized by its high diversity type and remarkable realism. As shown in Fig. \ref{fig:almpipeline}, a forger can easily choose any ALM model and generate different types of deepfake audio, such as speech, singing voice, music, and sound, for various forgery tasks.
These high-fidelity deepfake audio files are difficult for humans to discern, and the latest ALM models \cite{chen2024vall} have even achieved human parity in zero-shot text-to-speech (TTS) synthesis for the first time. This poses significant threats, including fraud, misleading public opinion, and privacy violations. Therefore, the urgent development of effective audio deepfake detection technologies is crucial.

As for countermeasures (CMs), the study of audio deepfake detection has been increasing. Many significant works have emerged \cite{tak2022automatic,xie2023learning,wang2024generalized,wang2024genuine,xie2023domain} around the ASVspoof competition series \cite{nautsch2021asvspoof,liu2023asvspoof}. In recent years, research in the audio deepfake detection (ADD) field has gradually shifted from improving in-distribution (ID) performance to focusing on generalization to wild domain. 
Tak et al. \cite{tak2022automatic} proposed a countermeasure (CM) based on self-supervised learning for front-end fine-tuning and AASIST for backend processing, which demonstrates remarkable generalization capabilities. Xie et al. \cite{xie2023learning} used three source domains for co-training and learned a self-supervised domain-invariant representations to improve the generalization of the CM. Wang et al. \cite{wang2024generalized} proposed a CM based on stable learning to address the issue of distribution shifts across different domain. Additionally, Wang et al. \cite{wang2024genuine} utilized masked autoencoder and proposed a reconstruction learning method focused on real data to enhance the model's ability to detect unknown forgery patterns. 
Despite these advancements, a crucial question remains:

\emph{Does current deepfake audio detection model effectively detect ALM-based deepfake audio? }

\begin{figure}[t]
	\centering
	\includegraphics[width=\linewidth]{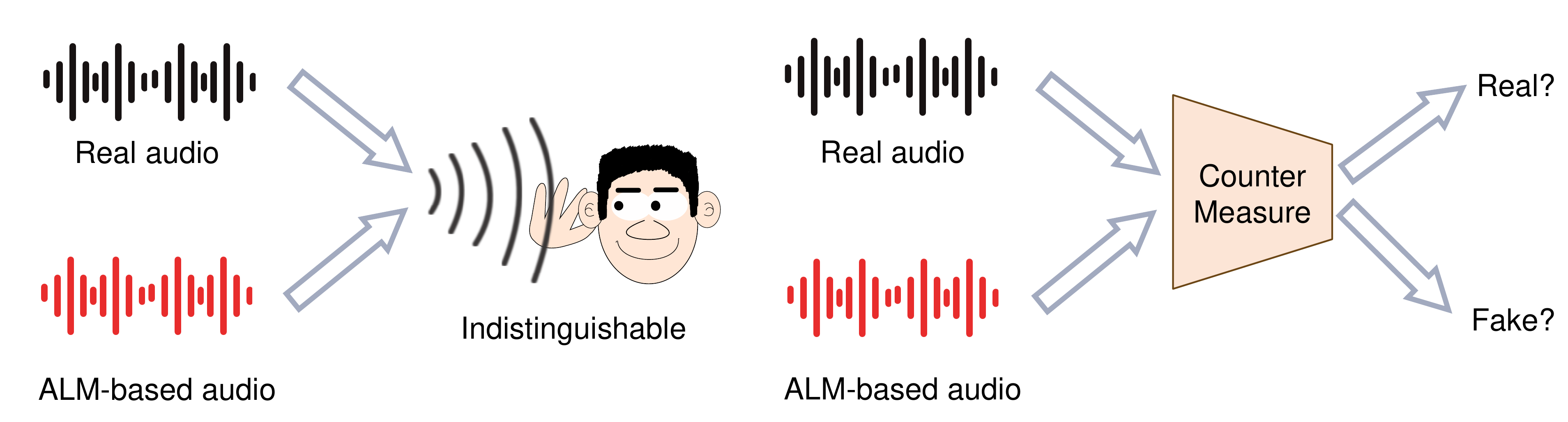}
	\caption{Does current countermeasure effectively detect ALM-based deepfake audio? }
	\label{fig:motivation}
\end{figure}


In this paper, we aim to address the aforementioned question by collecting as many ALM-based audio samples as possible and using the latest CM for deepfake detection. Specifically, we collected 12 types of ALM-based deepfake audio, which denote as A01-A12. Since most ALMs are not open-sourced, the majority of the audio samples have come from demo pages. The audio from demo pages is typically difficult for humans to distinguish as real or fake, posing a significant challenge for CM detection, as shown in Fig. \ref{fig:motivation}. For CM, the latest work on detecting ALM audio is Codecfake \cite{xie2024codecfake}, which focuses on the generation mechanism of ALM - neural codec, for deepfake detection. We takes advantage of Codecfake by training both traditional vocoder-trained CM and codec-trained CM to test ALM from A01 to A12. The experiments demonstrate that the codec-trained CM achieves the lowest average EER and 0\% EER under most ALM conditions, revealing that the current CM, specifically the codec-trained CM, can effectively detect ALM-based audio.

\begin{figure*}[t]
	\centering
	\includegraphics[width=6in]{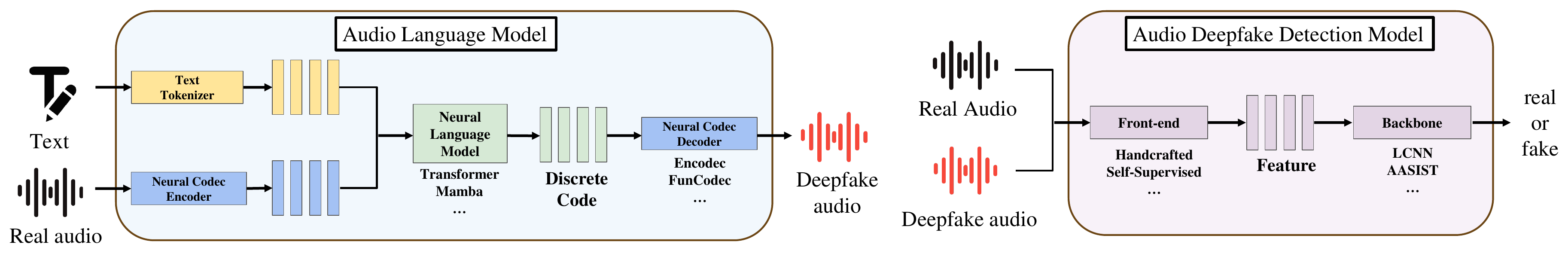}
	\caption{Generation and countermeasure pipeline for ALM-based deepfake audio.}
	\label{fig:Generation_pipeline}
\end{figure*}

\section{ALM-based deepfake audio}
\subsection{Generation pipeline}
ALM has rapidly developed due to advancements in both language model (LM) and neural codec model. In the left part of Fig. \ref{fig:Generation_pipeline}, it illustrates the pipeline used by most ALM models for generation. 
At first, the audio waveform is converted into discrete code representations through the encoder part of the neural codec. Then, the LM decoder performs contextual learning, where the discrete quantized tokens contain information about the speaker style. Finally, the non-autoregressive neural codec decoder generates the audio waveform from the discrete codes.

\subsection{Data collection}
We collected 12 type of ALM-based audio, denote as A01-A12. All collected audio samples can be found on the website\footnote{https://github.com/xieyuankun/ALM-ADD}. The numbering principle follows the publication date of the ALM works. Since most ALMs are not open-sourced, the majority of ALM-based audio samples come from demos.

\textbf{A01-AudioLM} \cite{borsos2023audiolm}.
AudioLM is a framework for high-quality audio generation with long-term consistency that maps the input audio to a sequence of discrete tokens and casts audio generation as a language modeling task in this representation space. We collected 48 fake speech samples and 28 real speech samples from the generation and continuation tasks on demo pages\footnote{https://google-research.github.io/seanet/audiolm/examples/}. Real speech comes from the ground truth or the speaker prompt, whereas AudioLM generated speech is categorized as fake speech. 

\textbf{A02-AudioLM-music} \cite{borsos2023audiolm}.
AudioLM is not limited to modeling speech. It can also learn to generate coherent piano music continuations. 4 pairs of music segments were collected from the piano continuation task on demo pages\footnote{https://google-research.github.io/seanet/audiolm/examples/}. Real audio comes from the 4-second piano prompts, whereas AudioLM generated music is categorized as fake audio.

\textbf{A03-VALL-E} \cite{wang2023neural}. VALL-E is a neural codec language model designed to generate discrete codes derived from EnCodec \cite{defossez2022high}, utilizing either textual or acoustic inputs. We used condition A1 from the Codecfake dataset \footnote{https://zenodo.org/records/11169781} for testing, which includes 4,451 real audio samples and 4,436 fake audio samples synthesized by VALL-E.

\textbf{A04-VALL-E X} \cite{zhang2023speak}. VALL-E X can generate high-quality audio in the target language with just a single utterance of the source language audio as a prompt. Similar to VALL-E, we used condition A2 from the Codecfake dataset for testing, which includes 4,451 real audio samples and 4,436 fake audio samples.

\textbf{A05-SpeechX} \cite{wang2023speechx}.
SpeechX is a versatile speech generation model leveraging audio and text prompts, which can deal with both clean and noisy speech inputs and perform zero-shot TTS and various tasks involving transforming the input speech. We collected 16 fake speech samples and 16 real speech samples from the TTS, content editing and target speaker extraction tasks on demo pages\footnote{https://www.microsoft.com/en-us/research/project/speechx/}. Real speech comes from the ground truth or the speaker prompt, whereas SpeechX generated speech samples are categorized as fake speech. 

\textbf{A06-UniAudio} \cite{yang2023uniaudio}.
UniAudio is a versatile audio generation model that conditions on multiple types of inputs and performs a variety of audio generation tasks. It treats all modalities as discrete tokens. We collected 30 fake speech samples and 18 real speech samples from the TTS tasks on demo pages\footnote{https://uniaudio666.github.io/demo\_UniAudio/}. Real speech comes from the ground truth or the speaker prompt, whereas Uniaudio generated speech is categorized as fake speech.

\textbf{A07-LauraGPT} \cite{chen2023lauragpt}.
LauraGPT can take both audio and text as input and output both modalities, and perform a broad spectrum of content-oriented and speech-signal-related tasks. We collected 8 fake speech samples and 8 real speech samples from the TTS tasks on demo pages\footnote{https://lauragpt.github.io/}. Real speech samples comes from the ground truth or the speaker prompt, whereas LauraGPT generated speech samples are categorized as fake speech. 

\textbf{A08-ELLA-V} \cite{song2024ella}.
ELLA-V is a simple but efficient LM-based zero-shot TTS framework, which enables fine-grained control over synthesized audio at the phoneme level. We collected 14 real speech samples and 32 fake speech samples from the generation and cloning tasks on demo pages\footnote{https://ereboas.github.io/ELLAV}. Real speech comes from the ground truth or the speaker prompt (speaker prompt-encodec), whereas ground truth Encodec regenerated and ELLA-V generated speech samples are categorized as fake speech. 

\textbf{A09-HAM-TTS} \cite{wang2024ham}.
HAM-TTS is a novel TTS system that leverages a hierarchical acoustic modeling approach. We collected 35 fake speech samples and 6 real speech samples from the TTS task on demo pages\footnote{https://anonymous.4open.science/w/ham-tts/}. Real speech samples come from the ground truth or the speaker prompt, whereas HAM-TTS generated speech samples are categorized as fake speech. 

\textbf{A10-RALL-E} \cite{xin2024rall}.
RALL-E is a robust language modeling method for text-to-speech synthesis that improves performance and reduces errors by using chain-of-thought prompting to decompose the task into simpler steps. We collected 5 fake speech samples and 5 real speech samples from the TTS tasks on demo pages\footnote{https://ralle-demo.github.io/RALL-E/}. Real speech comes from the ground truth, whereas RALL-E generated speech samples are categorized as fake speech. 

\textbf{A11-NaturalSpeech 3} \cite{ju2024naturalspeech}.
NaturalSpeech 3 is a TTS system that enhances speech quality, similarity, and prosody by using factorized diffusion models and factorized vector quantization to disentangle and generate speech attributes in a zero-shot manner. We collected 32 fake speech samples and 24 real speech samples from the TTS tasks on demo pages\footnote{https://speechresearch.github.io/naturalspeech3/}. Real speech samples comes from the ground truth or the speaker prompt, whereas NaturalSpeech 3 generated speech samples are categorized as fake speech. 

\textbf{A12-VALL-E 2} \cite{chen2024vall}.
VALL-E 2 is the latest advancement in neural codec language models that marks a milestone in zero-shot TTS, achieving human parity for the first time. We collected 91 fake speech samples and 33 real speech samples on demo pages\footnote{https://www.microsoft.com/en-us/research/project/vall-e-x/vall-e-2/}. Real speech samples come from the speaker prompt, whereas VALL-E 2 generated speech samples are categorized as fake speech.

\vspace{-2mm}
\section{Countermeasure}

For the countermeasure, we consider two aspects: the training datasets and the audio deepfake detection (ADD) model. For the dataset, we selected models trained on the classic vocoder-based deepfake dataset ASVspoof2019LA and the latest codec-based deepfake dataset Codecfake. This approach allows us to verify whether countermeasures trained on traditional vocoder datasets can effectively detect ALM-based audio, as well as to evaluate the effectiveness of CM trained on the Codecfake dataset in practical ALM in the wild tests. 

As the feature of ADD model, we selected handcrafted features Mel-spectrogram and pre-trained features wav2vec2-xls-r \cite{oneata2023towards}. For the Mel-spectrogram, we used 80-dimensional Mel-spectrograms to match the features of most conventional audio generation tasks. For wav2vec2-xls-r, we froze the weights and used its fifth hidden layer as the feature, due to its proven superiority in previous research \cite{lee22q_interspeech}. For the backbone networks, we chose LCNN \cite{lavrentyeva2019stc} and AASIST \cite{jung2022aasist}, which are currently the most commonly used backbone networks in the filed of ADD. The CM pipeline is shown in Fig. \ref{fig:Generation_pipeline}.

\section{Experiments}
\subsection{Experiments setting}
In the experiments, we trained on two different datasets. The vocoder-trained CM was trained using the ASVspoof2019 LA (19LA) training set, which includes 25,380 training samples and 24,844 validation samples. There are six spoofing methods in total, and the validation set was used only to select the best-performing model without participating in the training.

The codec-trained CM was trained using the Codecfake training set and used the validation set for model selection. The Codecfake training set contains 740,747 samples, and the validation set contains 92,596 samples, with a total of six codec reconstruction methods.

For the test set, we first conducted preliminary performance tests on the CM using the 19LA test set and an in-the-wild (ITW) dataset. The 19LA test set includes 71,237 audio samples with attack types (A07-A19) that are not seen in the training set. ITW dataset includes 31,779 audio samples. Both real and fake audio samples are collected from publicly available sources, such as social networks and video streaming platforms, which may contain background noise. This dataset is intended to evaluate the generalizability of detection models, including cross-dataset evaluation.

\subsection{Implementation details}
In the pre-processing stage for the countermeasure models (CMs), all audio samples were initially down-sampled to 16,000 Hz and adjusted to a uniform duration of 4 seconds through trimming or padding. For the mel-spectrogram feature extraction, we derived an 80-dimensional mel-spectrogram. For the self-supervised feature extraction, we utilized the Wav2Vec-XLS-R model\footnote{https://huggingface.co/facebook/wav2vec2-xls-r-300m} with frozen parameters, extracting 1024-dimensional hidden states as feature representations.

All CMs were trained using the Adam optimizer with a learning rate of 5$\times10^{-4}$. The vocoder-trained CM underwent 100 epochs of training using a weighted cross-entropy loss, assigning a weight of 10 to the real class and 1 to the fake class. The learning rate was halved every 10 epochs. In contrast, the codec-trained CM was trained for 10 epochs, with the learning rate halved every 2 epochs. The model showing the best performance on the validation set was chosen for evaluation.

For experimental evaluation, we used the official implementation for calculating EER\footnote{https://github.com/asvspoof-challenge/2021/blob/main/eval-package/eval\_metrics.py}, maintaining precision to three decimal places. To compute the confusion matrix, we used a threshold of 0.5 to distinguish between real and fake predictions, utilizing calculations from SKlearn.

\vspace{-2mm}
\section{Results and Discussion}
\subsection{Results on vocoder-based dataset}
To verify the generalization ability of the CM, we first test the performance on traditional vocoder-based dataset as shown in the left part (19LA, ITW) of Table \ref{tab:vocoder-trained} and Table \ref{tab:codec-trained}. Specifically, the vocoder-trained W2V2-AASIST achieve the best equal error rate (EER) of 0.122\% in 19LA test set and 23.713\% in ITW dataset. Especially in 19LA, to the best of our knowledge, our CM can achieve the lowest EER, indicating the generalizability of the CM. Additionally, in cross-domain training scenarios, where only Codecfake is used for training and testing is conducted on 19LA and ITW, W2V2-AASIST also achieved good results, with an EER of 3.806\% on 19LA and 9.606\% on ITW. 

\vspace{-0.5mm}
\subsection{Results on ALM-based dataset}
\begin{figure*}[!t]
	\centering
	\subfloat{\includegraphics[width=5in]{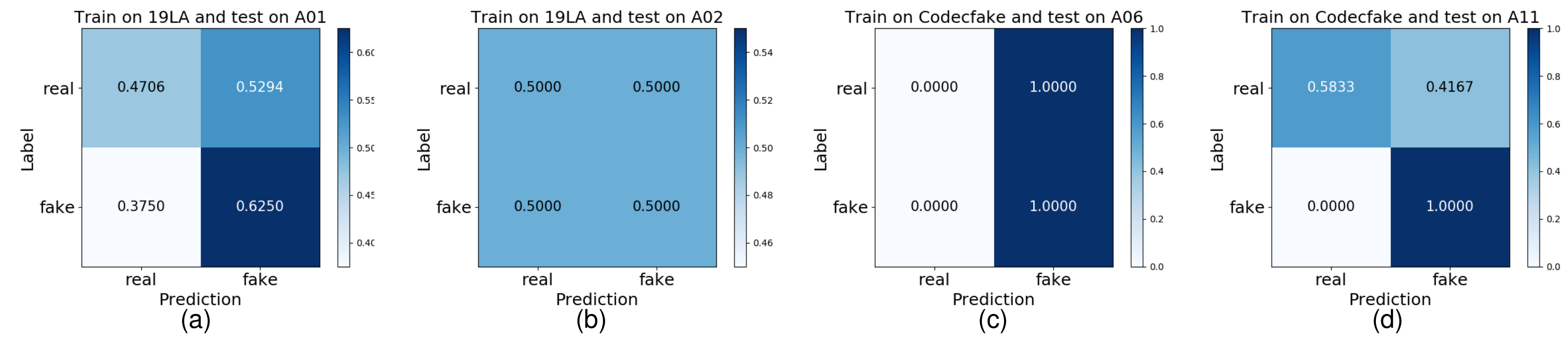}}
	\hfil
	\caption{The confusion matrices under different test conditions. (a), (b) correspond to W2V2-AASIST trained on the 19LA training set and tested on A01 and A02. (c), (d) correspond to W2V2-AASIST trained on the Codecfake training set and tested on A06 and A11.}
	\label{fig:confusion} 
\end{figure*}

We tested the collected and generated ALM data. For the 19LA-trained CM, the overall average (AVG) result was not very good, with the lowest AVG being 24.403\% for W2V2-AASIST. The EER performance is consistent with the CM's performance and  vocoder-based data. For instance, Mel-LCNN performed poorly on 19LA and ITW, and similarly, its performance on ALM was also not good, with an AVG as high as 39.592\%. Furthermore, we conducted a confusion matrix analysis for the two worst cases, A01 and A02, of W2V2-AASIST, as shown in Fig. \ref{fig:confusion}(a) and \ref{fig:confusion}(b). Fig. \ref{fig:confusion}(a) corresponding to an EER of A01 for 39.435\%, it can be seen that 52.94\% of real speech samples are misclassified as fake, and 37.50\% of fake speech samples are misclassified as real. The CM fails to detect ALM-based audio and produces false positives on genuine audio. Fig. \ref{fig:confusion}(b) shows a similar situation, with 50\% of genuine audio misclassified as fake and 50\% of fake audio misclassified as genuine.

For the codec-trained CM, the results are very surprising. Most EER values are 0.000\% when rounded to three decimal places, indicating a very high distinction between real and fake audio. The best-performing model is still W2V2-AASIST, achieving an EER of 9.424\%. This also demonstrates the model's stability in performance across different training datasets. Furthermore, we attempted a detailed analysis of the worst cases, as shown in Fig. \ref{fig:confusion}(c) and (d). It can be seen that the poor performance is due to real speech samples being misclassified as fake, while fake samples are correctly identified. Specifically, in A06, 100\% of the real speech was classified as fake, and in A11, 41.67\% of the real speech was classified as fake. This indicates that the codec-trained CM can recognize ALM-based audio but lacks generalization to real data from other domains. Rather than speech, the situation for music in A02 is also very poor. For the same reason, the model classified all samples as fake. These findings suggest that when the CM encounters new audio types, new features and backbones need to be considered to adapt to these new types of audio.

\begin{table*}[t]
	\caption{EER (\%) results for CM trained by 19LA training set. AVG represents the average EER across A01-A12.}
	\label{tab:vocoder-trained}
	\centering
	\setlength{\tabcolsep}{0.4mm}
	\renewcommand{\arraystretch}{1.1}
	\begin{tabular}{|c|c|c|c|c|c|c|c|c|c|c|c|c|c|c|c|c|}
		\hline
		Feature & Backbone & 19LA & ITW & A01 & A02 & A03 & A04 & A05 & A06 & A07 & A08 & A09 & A10 & A11 & A12 & AVG \\
		\hline
		Mel & LCNN & \cellcolor{gray!5}5.084 & \cellcolor{gray!52}47.021 & \cellcolor{gray!53}46.131 & \cellcolor{gray!57}50.000 & \cellcolor{gray!24}21.233 & \cellcolor{gray!4}3.241 & \cellcolor{gray!33}31.250 & \cellcolor{gray!51}43.858 & \cellcolor{gray!44}37.500 & \cellcolor{gray!44}37.500 & \cellcolor{gray!56}49.286 & \cellcolor{gray!45}40.000 & \cellcolor{gray!38}33.854 & \cellcolor{gray!51}45.255 & \cellcolor{gray!45}39.592 \\
		\hline
		W2V2 & LCNN & \cellcolor{gray!1}0.625 & \cellcolor{gray!45}41.385 & \cellcolor{gray!50}43.304 & \cellcolor{gray!57}50.000 & \cellcolor{gray!1}0.608 & \cellcolor{gray!3}2.093 & \cellcolor{gray!7}6.250 & \cellcolor{gray!38}33.333 & \cellcolor{gray!44}37.500 & \cellcolor{gray!35}30.208 & \cellcolor{gray!16}14.048 & \cellcolor{gray!45}40.000 & \cellcolor{gray!38}33.854 & \cellcolor{gray!45}39.477 & \cellcolor{gray!31}27.556 \\
		\hline
		W2V2 & AASIST & \cellcolor{gray!0}0.122 & \cellcolor{gray!23}23.713 & \cellcolor{gray!46}39.435 & \cellcolor{gray!57}50.000 & \cellcolor{gray!0}0.225 & \cellcolor{gray!1}0.833 & \cellcolor{gray!7}6.250 & \cellcolor{gray!38}33.333 & \cellcolor{gray!44}37.500 & \cellcolor{gray!28}25.000 & \cellcolor{gray!19}16.905 & \cellcolor{gray!23}20.000 & \cellcolor{gray!34}30.208 & \cellcolor{gray!38}33.150 & \cellcolor{gray!24}24.403 \\
		\hline
	\end{tabular}
\end{table*}

\begin{table*}[t]
	\caption{EER (\%) results for CM trained by Codecfake training set. AVG represents the average EER across A01-A12.}
	\label{tab:codec-trained}
	\centering
	\setlength{\tabcolsep}{0.4mm}
	\renewcommand{\arraystretch}{1.1}
	\begin{tabular}{|c|c|c|c|c|c|c|c|c|c|c|c|c|c|c|c|c|}
		\hline
		Feature & Backbone & 19LA & ITW & A01 & A02 & A03 & A04 & A05 & A06 & A07 & A08 & A09 & A10 & A11 & A12 & AVG \\
		\hline
		Mel & LCNN & \cellcolor{gray!27}26.826 & \cellcolor{gray!43}42.635 & \cellcolor{gray!31}31.696 & \cellcolor{gray!57}50.000 & \cellcolor{gray!7}7.393 & \cellcolor{gray!9}9.553 & \cellcolor{gray!44}43.750 & \cellcolor{gray!46}45.556 & \cellcolor{gray!12}12.500 & \cellcolor{gray!21}21.652 & \cellcolor{gray!0}0.000 & \cellcolor{gray!45}40.000 & \cellcolor{gray!38}33.854 & \cellcolor{gray!45}44.705 & \cellcolor{gray!28}28.388 \\
		\hline
		W2V2 & LCNN & \cellcolor{gray!4}4.433 & \cellcolor{gray!5}4.975 & \cellcolor{gray!17}17.262 & \cellcolor{gray!57}50.000 & \cellcolor{gray!0}0.450 & \cellcolor{gray!1}1.080 & \cellcolor{gray!0}0.000 & \cellcolor{gray!39}39.444 & \cellcolor{gray!0}0.000 & \cellcolor{gray!1}1.562 & \cellcolor{gray!0}0.000 & \cellcolor{gray!0}0.000 & \cellcolor{gray!4}3.646 & \cellcolor{gray!3}3.164 & \cellcolor{gray!9}9.717 \\
		\hline
		W2V2 & AASIST & \cellcolor{gray!4}3.806 & \cellcolor{gray!9}9.606 & \cellcolor{gray!4}3.869 & \cellcolor{gray!57}50.000 & \cellcolor{gray!0}0.225 & \cellcolor{gray!0}0.135 & \cellcolor{gray!0}0.000 & \cellcolor{gray!57}50.000 & \cellcolor{gray!0}0.000 & \cellcolor{gray!0}0.000 & \cellcolor{gray!0}0.000 & \cellcolor{gray!0}0.000 & \cellcolor{gray!9}8.854 & \cellcolor{gray!0}0.000 & \cellcolor{gray!9}9.424 \\
		\hline
	\end{tabular}
\end{table*}

\subsection{Discussion}
\vspace{-2mm}
Some codec-trained CMs still exhibit shortcomings, particularly high false negative (FN) rates, which suggest several recommendations for future research on countering ALM-based audio. At first, the Codecfake dataset may lacks generalization to other audio types such as music and sound. Even though codec-trained CMs can detect ALM-based audio, they can also misclassify genuine audio. From dataset perspective, this may necessitate enriching the codec with a variety of audio types. As for CM, specialized features for different audio types rather than relying solely on speech self-supervised feature such as W2V2 may be needed. 
Additionally, the high FN rates indicate deficiencies in the classifier's ability to learn from real-world audio data. This is apparent due to the limited diversity of real-world domains covered by a single dataset during training, and the inadequate representation of real audio in pre-trained features. Therefore, strategies such as co-training with supplementary real audio datasets or enhancing pre-trained features could significantly enhance performance. 
\vspace{-2mm}
\section{Conclusions}
In this paper, we attempt to address a novel question: does the current deepfake audio detection model effectively detect ALM-based deepfake audio? We evaluate this by collecting and generating the latest 12 types of ALM-based audio and assessing them using SOTA performance CM. The surprising results indicate that codec-trained CMs can effectively detect these ALM-based audios, with most EERs approaching 0\%. This indicates that the current CM, specifically the codec-trained CM trained with Codecfake dataset, can effectively detect ALM-based audio.
\vspace{-2mm}
\section{Acknowledgements}
This work is supported by the National Natural Science
Foundation of China (NSFC) (No.62101553, No.62306316,
No.U21B20210, No. 62201571).

\bibliographystyle{IEEEtran}

\bibliography{mybib}


\end{document}